# Hot-Ion Modes in Globus-M2 and Saturation of Energy Confinement Time Scaling in Spherical Tokamaks with Toroidal Magnetic Field of 1 T and Above.

*G.S. Kurskiev, V.B. Minaev, N.V. Sakharov, V.K. Gusev, Yu.V. Petrov, I.V. Miroshnikov, F.V. Chernyshev, N.N. Bakharev, E.O. Kiselev, A.Yu. Telnova, E.E. Tkachenko, N.S. Zhiltsov and Globus- M2 Team*.*

*Ioffe Institute, St. Petersburg, Russia*

* See G.S. Kurskiev et al 2024 (http://dx.doi.org/10.1063/5.0211866) for the Globus-M Team.

gleb.kurskiev@mail.ioffe.ru,

gleb.kurskiev@gmail.com

## Abstract.

In a small spherical tokamak with minor radius of 0.22 m and toroidal magnetic field of 1 T, it is possible to heat ions of a sufficiently dense plasma to an extremely high temperature up to 50 million Kelvin. To do this, it is necessary to transfer a sufficiently large torque to stabilize ion-scale turbulence and achieve ion heat transport at neoclassical level reaching extremely low values of plasma collisionality. It is also necessary to ensure good thermal insulation of electrons, which is always determined by turbulent transport. In a spherical tokamak, the toroidal magnetic field has a strong beneficial effect on suppressing turbulent electron heat fluxes. In the Globus-M2 tokamak, when heating plasma with high-energy atomic beams, a significant improvement in the plasma thermal energy confinement is observed with an increase in the toroidal magnetic field from 0.5 to 1 T. The comparison of our results with experiments on ST40 tokamak operating with toroidal magnetic field of 2 T indicates that further strong improvement of thermal insulation of spherical tokamak plasma in the region of higher magnetic fields is not expected.

## *Introduction.*

Spherical tokamaks (ST), or tokamaks with small aspect ratio A=R/a<2 (R and a - major and minor radii correspondently), provide stable operation with high plasma current at a relatively low toroidal magnetic field ($B_T$), reaching a large toroidal beta $\beta_T \geq 30\%$ [1] and normalized beta $\beta_N \geq 6$ [2] values. Despite the achievement of the high confinement mode (H-mode) regime, where the ion heat transport remained neoclassical level [3], the first experiments on STs with $B_T \leq 0.5$ T showed a rather poor thermal insulation of the plasma with, with H-factor (the ratio of energy confinement time to predictions of widely used IPB98(y,2) scaling of approximately 0.7 [4,5]. It was found that the main loss channel was anomalous electron heat transport caused by micro-tearing (MTM) and electron temperature gradient (ETG) modes, and it rapidly decreased with decreasing collisionality [6-8], causing a strong dependence of the energy confinement time on toroidal magnetic field in STs for $B_T$ up to 0.8 T [9-11]. Various independent research carried out on different STs have shown that energy confinement time ($\tau_E$) exhibits strong dependence on $B_T$ and a moderate dependence on $I_p$ [3].

However, there were reasonable doubts concerning the achievability of good confinement in STs at low collisionality $\nu^* \propto n/T^2$ expected for operational regimes with high $B_T$. For such conditions the development of ion scale ion temperature gradient mode (ITG) that drives high ion heat fluxes was suggested theoretically long time ago [9]. Moreover, on NSTX for regimes with low electron collionality $\nu_e^* < 0.15$ the strong enhancement of ion heat transport above the neoclassical level was observed [10] in discharges with $B_T$=0.55 T. The first experiments on plasma heating using neutral beam in Globus-M2 at $B_T$=0.8 T and collisionality <0.015 also demonstrated poor ion heating due to the development of the ITG mode. Nevertheless, later experiments on the Globus-M2 ($B_T$=0.9) and ST40 ($B_T$=1.9) tokamaks demonstrated the achievement of the hot ion mode with fusion-relevant ion temperatures of 4 and 9 keV correspondently using similar plasma heating scenarios with two neutral beams [12,13]. The obtained result inspires a certain optimism regarding ambitious projects based on STs using D-T reaction (the STEP project [14]), p-$B^{11}$ fusion (EHL-2 [15] project) and a prototype of a fusion neutron source for rector materials testing [16]. The design of any of these facilities requires reliable predictions of the expected plasma parameters far beyond the explored area. How far can we go, based on the results achieved so far? The only way to solve this problem is to consistently build experimental setups with parameters approaching the cherished goal.

This paper provides the review of the main energy confinement features of spherical tokamak plasmas heated by the neutral beams in a wide range of $B_T$=0.5-0.9 T discovered at the Globus-M2. The paper is organized as follows. The first section presents the typical hot ion mode scenario in Globus-M2 using NBI at the current-ramp-up phase. The second section describes the influence of $I_p$, $B_T$ and $\nu^*$ on energy confinement during the quasi-steady state phase of the discharge. The third is devoted to the influence of the injected atoms mass on ion heating and ion heat transport. A comparison of the plasma energy confinement in Globus-M2 and ST40 tokamak is presented in the discussion section.

## ***Ion heating during plasma current ramp-up phase.***

The simplest way to obtain high ion temperatures in Globus-M2 is early neutral beam injection (NBI) at the plasma current-ramp-up phase. Such plasma heating scenario previously allowed to suppress turbulence in the plasma core by the weak magnetic shear and achieve peaked density profiles [17,18] on Globus-M with $B_T \leq 0.5$ T. This mode was limited by the resonant magnetic surface $q$=1 ($q$ – local value of the safety factor) formation that led to start of sawtooth oscillations destructing the good confinement in the plasma core. Increasing $B_T$ up to 0,9 T one could expect reaching regimes with high ion temperatures due to significant reduction of neoclassical ion heat diffusivity at higher $B_T$ and plasma current.

Figure (Figure 1) illustrates the typical Globus-M2 plasma discharge with $B_T$=0.9 T with early NBI at the current ramp-up. The deuterium beam with energy of 45 keV and power of 0.9 MW is injected to the deuterium plasma. Plasma current reaches its maximum of 0.4 MA at 160 ms, 25 ms before the sawtooth oscillations starts. At the same time one can see the typical signs of transition to high confinement mode (H-mode): density rise, drop of $D_\alpha$ emission and development of ELMs that indicates the increase of the pressure gradient at the plasma edge. Unlike the electron temperature in the plasma center that monotonically rise to 1.5 keV and doesn't change significantly further the ion temperature behaves more complicated. The rapid rise of the $T_i^0$ (ion temperature in the vicinity of the magnetic axis) up to 4 keV, i.e. the plasma is in the hot-ion-mode with $T_i/T_e$ ratio higher than 2.6. It is worth noting that density profile is also peaked, and local density in the plasma core exceeds $10^{20}$ $m^{-3}$ (see Figure 2). When the Alfven velocity decreases to the velocity of the fast ions due to the density rise the excitation of toroidal Alfven eigenmodes are observed on MHD signal. This has a negative impact on plasma performance. We observe that the rise in $T_i^0$ is replaced by a decrease. This may be due to the fact that the fast ions losses caused by TAE decrease plasma heating power as well as the torque applied to the plasma. The latter leads to a decrease in the speed of toroidal rotation of the plasma, which is a key parameter affecting the thermal insulation of plasma ions. The plasma edge is also affected - a back transition to the low confinement mode (L-mode) occurs. The subsequent inevitable formation of a resonant magnetic surface q=1 in the plasma central region due to poloidal magnetic flux diffusion leads to internal kink mode development with poloidal and toroidal mode numbers m=1/n=1. It finally destroys the peaked ion temperature and density profiles; however, the discharge continues further with regular sawtooth oscillations and flat temperature and density profiles inside mixing region.

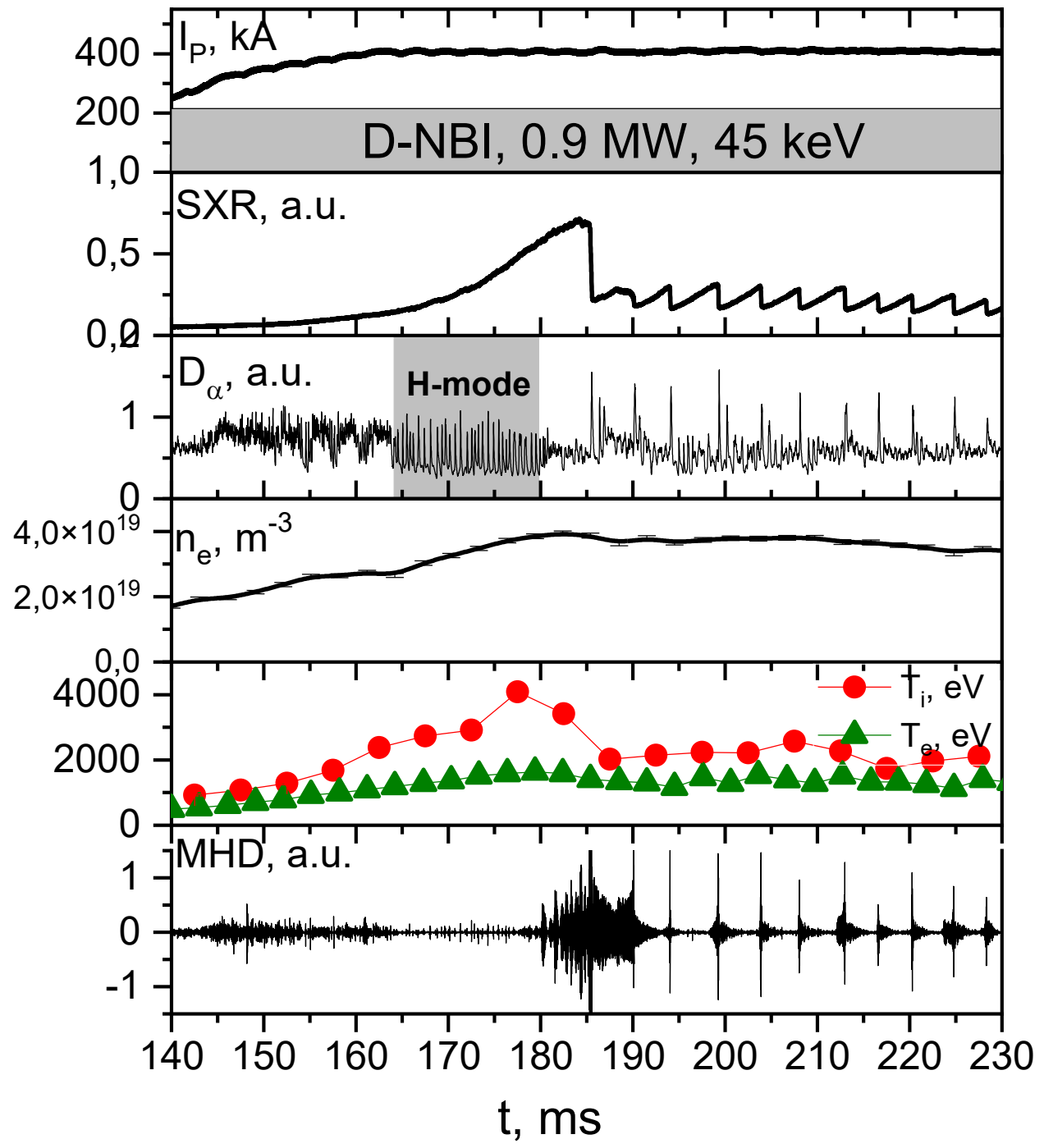

*Figure 1. Time traces of typical Globus-M2 with $B_T$=0.9 T plasma discharge with early NBI at the current ramp-up. From top to bottom plasma current, $D_\alpha$ emission, line averaged density, core electron (green triangles) and ion (red circles) temperatures, Mironov signal.*

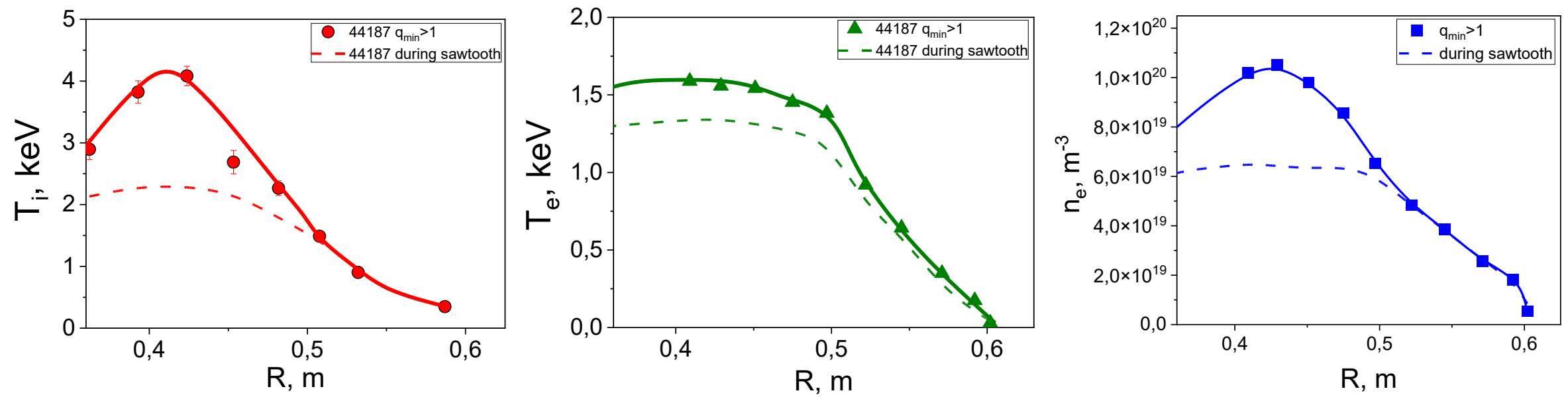

a) b) c)

*Figure 2. Spatial distributions of ion temperature (a), electron temperature (b), and electron density (c) immediately before the onset of sawtooth oscillations. The dotted lines show the corresponding profiles during sawtooth oscillations.*

## *Effect of $I_p$ and $B_T$ on $\tau_E$*

The evolution of plasma energy confinement during the quasi-stationary phase of discharges with sawtooth oscillations has been analyzed. A total of 300 discharges from the Globus-M2 database were selected, with the main parameters varying in the following ranges: $I_p$=0.15 – 0.4 MA, $B_T$ = 0.5-0.95 T, $n_e$=1.5-9 $10^{19}$ м$^{-3}$, $P_{NBI}$ = 0.3-0.9 MW. One of the primary diagnostic tools was Thomson scattering diagnostics. The measurement frequency of 330 Hz provides 10–15 profile samples during the quasi-stationary phase of a single discharge. In combination with magnetic reconstruction data, this allows to estimate electron stored energy ($W_e$). Figure 3 shows linear regression results for $W_e$ and the total plasma stored energy ($W_{MHD}$, derived using the Shafranov integral method) versus key engineering parameters. The magnetic and kinetic energy estimates are in a good agreement. Both $W_{MHD}$ and $W_e$ show a stronger dependence on plasma current than on toroidal magnetic field, suggesting that $\tau_E$ behaves similarly. Simultaneous increases in $I_p$ and $B_T$ lead to a significant rise in electron temperature and total plasma stored energy, and thus in the energy confinement time.

For clarity we compare the energy confinement for two discharges with different plasma current and magnetic field: $I_p$ = 0.3 MA, $B_T$ = 0.7 T and $I_p$ = 0.4 MA, $B_T$ = 0.9 T a fixed plasma density of 5 × $10^{19}$ m$^{-3}$, where the highest $T_i$ values are observed. A simultaneous 30% increase in $I_p$ and $B_T$ resulted in a notable rise in both electron and ion temperatures, as measured by Thomson scattering diagnostics and charge-exchange recombination spectroscopy (see Figure 4). In both discharges, the plasma was in a hot-ion mode with $T_i \approx 1.5\,T_e$. For both discharges, the NBI power was 0.7 MW, of which ~0.4 MW was deposited into the plasma. Fast particle orbit losses were relatively low (≤15% of injected power), while charge exchange losses were dominant and could reach up to 35%. Ohmic heating power was the same for both cases $P_{OH}$ ~0.42 MW, due to the rise of the electron temperature. Since the deposited heating power was the same in both cases, the increase in plasma temperature is attributed to a 1.5-fold increase in energy confinement time (from 7.5 ± 1.4 ms to 11.5 ± 2.5 ms), that is consistent with previously derived Globus-M2 scaling laws: $\tau_E^{Globus-21} \sim Ip^{0.53} \cdot Bt^{1.05}$ , $\tau_E^{GLB_2020} \sim Ip^{0.43} \cdot Bt^{1.19}$ and slightly higher than the IPB98(y,2) scaling predictions: $\tau_E^{IPB98(y,2)} \sim Ip^{0.93} \cdot Bt^{0.15}$. In contrast, the ion energy confinement time remained ~22 ms for both cases, corresponding to an effective ion heat diffusivity of ~1 m$^2$/s. Thus, overall plasma energy confinement is predominantly governed by electron heat transport. The electron energy confinement time $\tau_E^e$, being lower than $\tau_E$, increased from 4.0 ms to 7.5 ms as $B_T$ rose from 0.7 T to 0.9 T and $I_p$ from 0.3 MA to 0.4 MA.

Figure 5 presents the comparison of the normalized inverse electron temperature gradient scale length ($R/L_{Te}$) with threshold values the ETG and TEM instabilities at r/a = 0.7 estimated using approximations from [19,20]. Most data points lie below the TEM threshold, suggesting a low probability of TEM-driven turbulence in these regimes. In contrast, ETG thresholds are significantly exceeded across the dataset. It also should be noted that, gyrokinetic simulations using the GENE code for Globus-M2 consistently predict strong ETG growth rates, peaking at k ≈ 100 cm$^{-1}$.

The dimensionless energy confinement time $\Omega\tau_E \sim B_T\tau_E$ has a moderate dependence on effective collisionality $\nu_{eff} \sim \left(n_e Z_{eff}\right) / T_e{}^2$: $B_T\tau_E \sim \nu_{eff}{}^{-0,45}$, that is in a good agreement with dependence previously carried out on Globus-M for $B_T$≤0.5 T [21]. Analysis of a larger dataset does not confirm the increase in absolute value of the colliosnality exponent that we observed comparing the first experiments on Globus-M2 with Globus-M data [11]. The difference in the collisionality exponents for Globus-M/M2 ($|\alpha_\nu|$=0.4-0.45) and NSTX/MAST experiments ($|\alpha_v|$=0.8-1.2) is apparently a consequence of differences in the tokamak sizes and normalized Larmor radius.

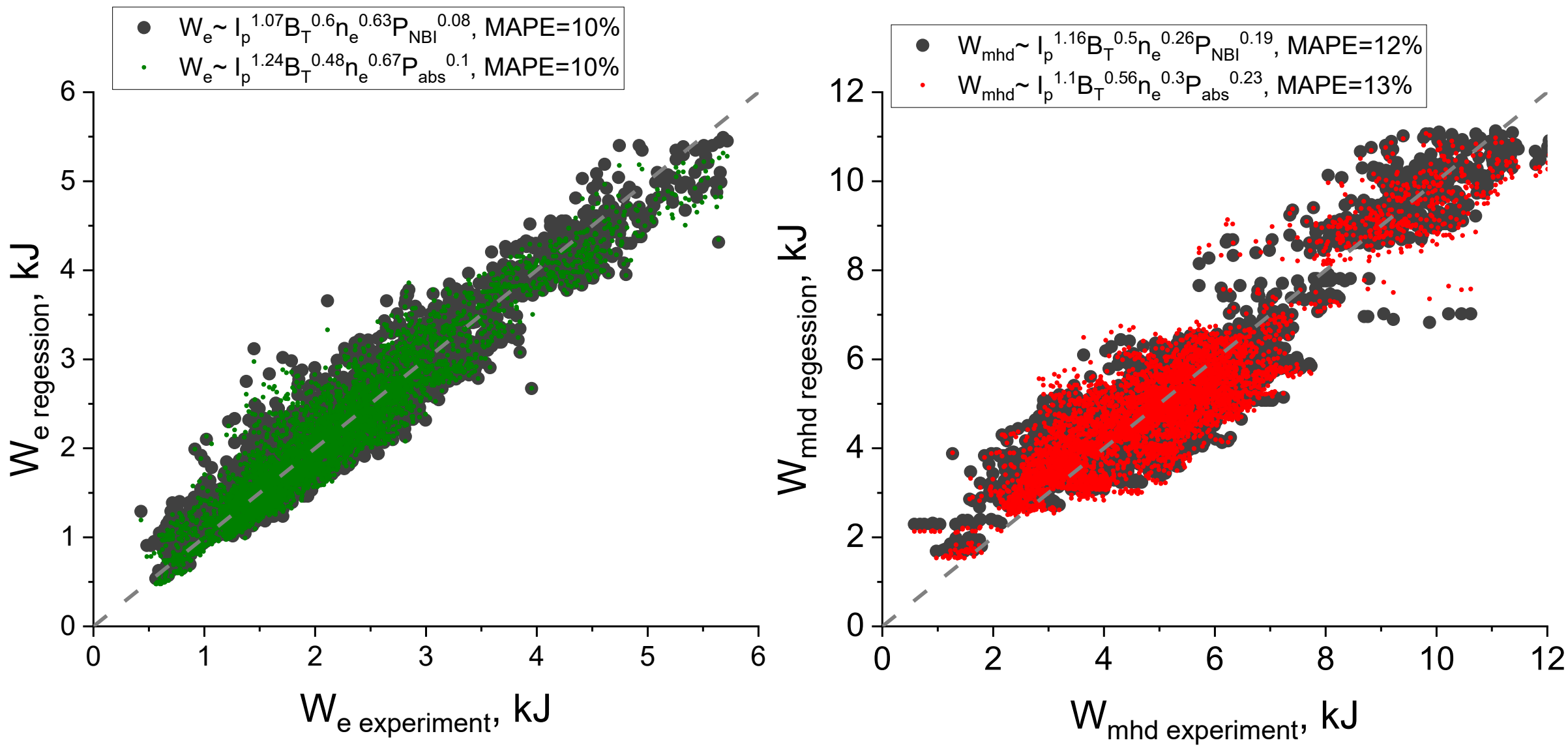

*Figure 1. Results of linear regression for (a) electron stored energy from Thomson scattering diagnostics and (b) total plasma stored energy from magnetic measurements, as a functions of plasma current, toroidal magnetic field, electron density, and auxiliary heating power.*

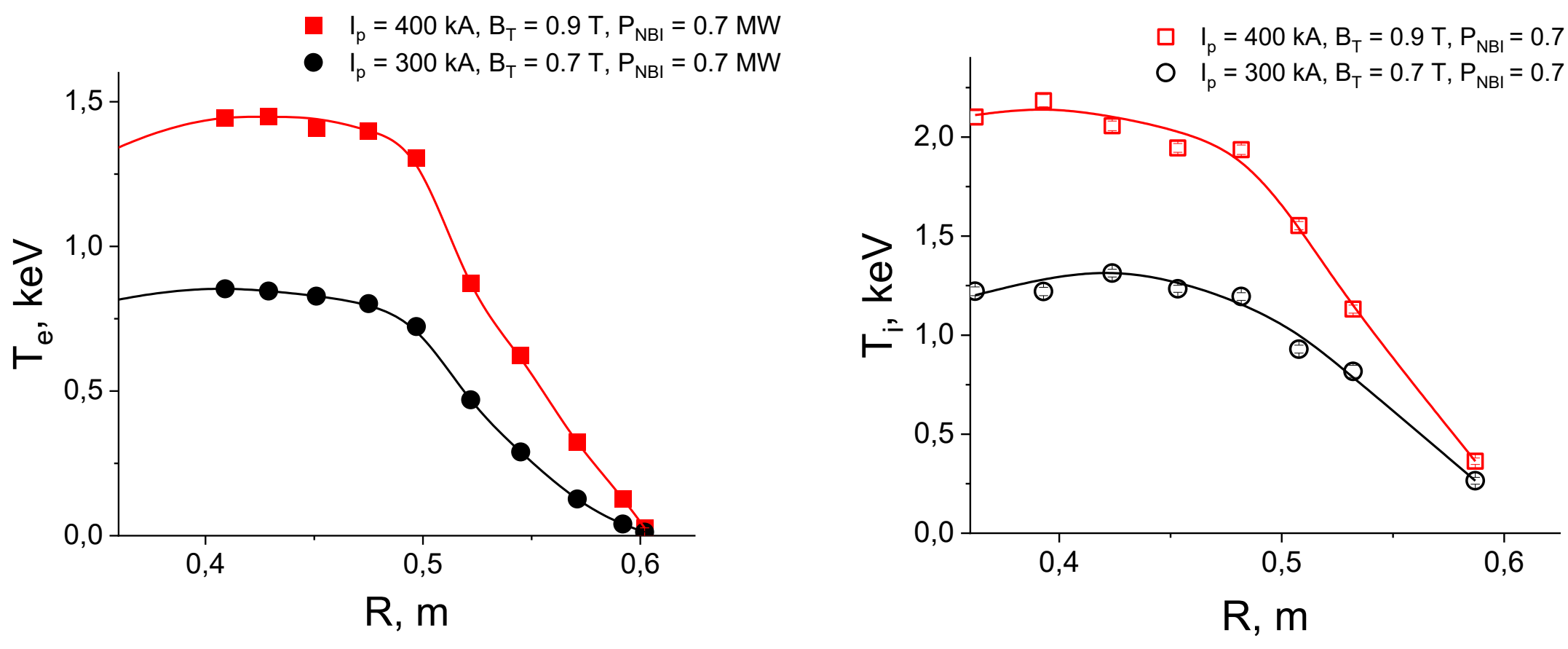

a) б)

*Figure 2. Experimental profiles of (a) electron temperature and (b) ion temperature for discharges with $I_p$ = 300 кA, $B_T$ =0.7 T and $I_p$ = 400 кA, $B_T$ =0.9 T, at fixed $P_{NBI}$.*

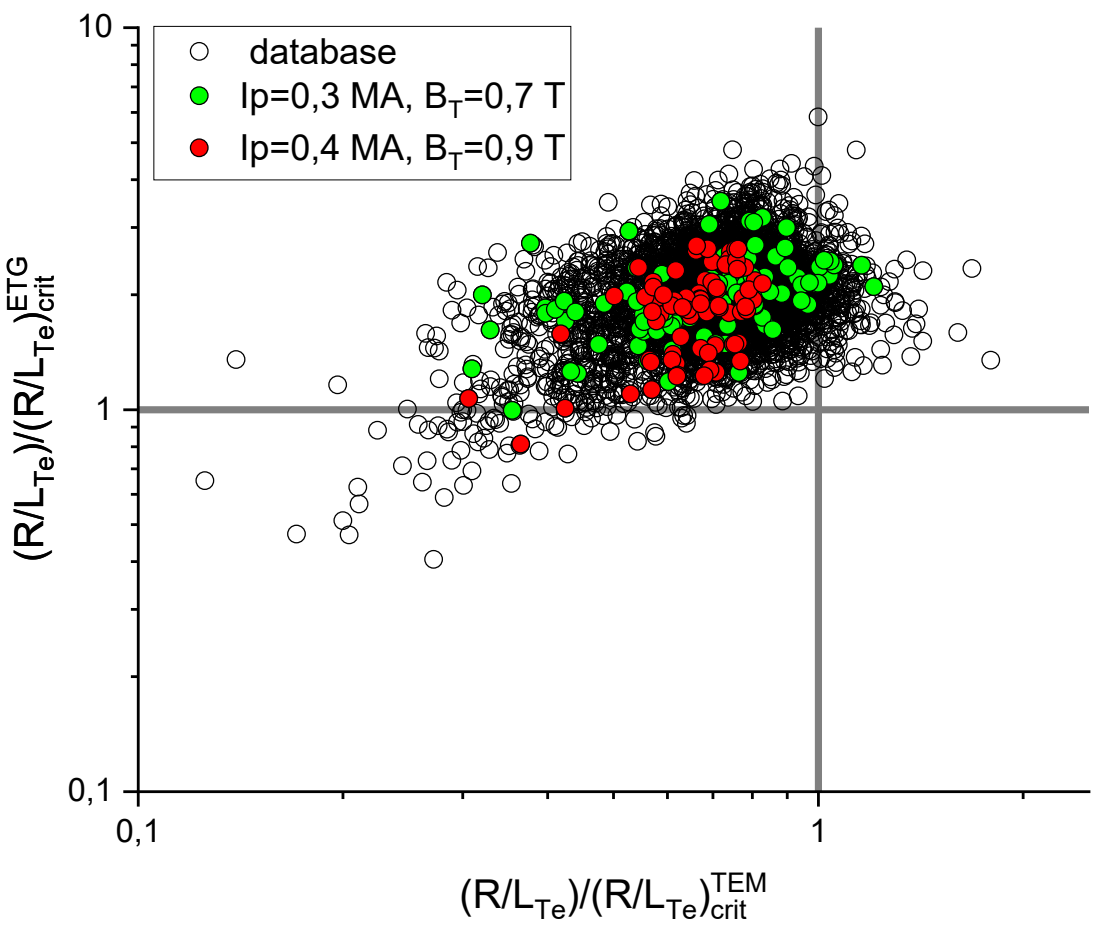

*Figure 3. Comparison of the normalized inverse electron temperature gradient scale length ($R/L_{Te}$) at r/a = 0.7 with threshold values for the onset of ETG and CTEM instabilities, based on the Globus-M2 dataset shown in Figure 3. Colored points indicate experimental data for $n_e = 5 \times 10^{19}$ $m^{-3}$ and $P_{NBI}$ = 0.7 MW.*

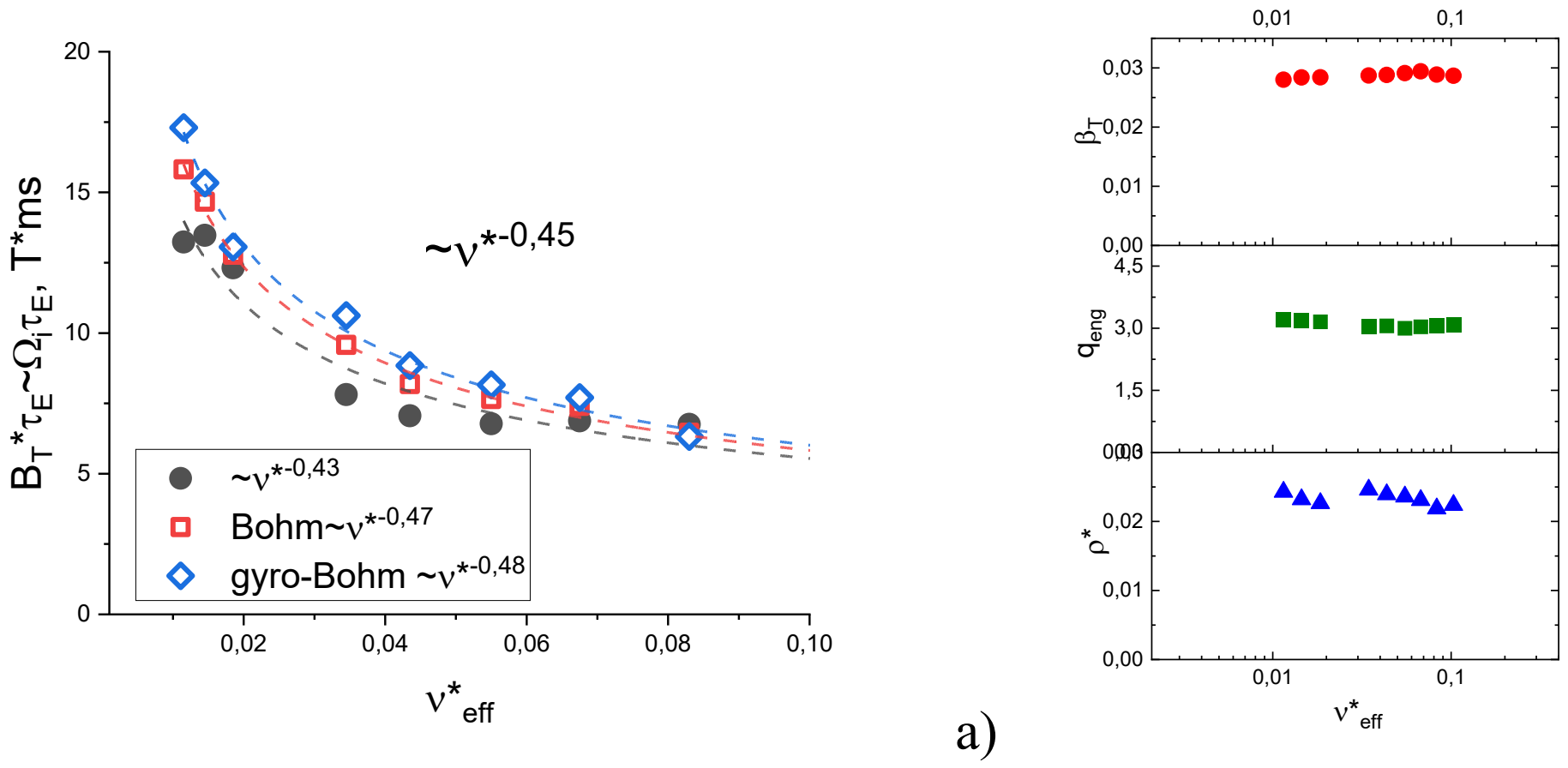

*Figure 4. Dependence of the normalized energy confinement time on plasma collisionality (a) at fixed values of safety factor, beta, and normalized ion Larmor radius (b).*

## ***Effect of injected ion mass on ion heating.***

Although sawtooth oscillations tend to flatten both density and ion temperature profiles, they are not a fundamental barrier to achieving fusion-relevant temperatures in spherical tokamaks. When using two neutral beam injectors (NBI-1 and NBI-2) for plasma heating, high ion temperatures can still be achieved during the stationary phase of the discharge, even in the presence of sawtooth activity. For this we use a two-staged NBI heating scheme: one injector is switched on during the plasma current ramp-up phase and remains active throughout the discharge, while the second starts at the current flat-top phase. Utilizing this approach, the central ion temperatures comparable to those achieved in $q_{min} > 1$ regimes were routinely obtained (see Figure 7a).

It was also found that deuterium beams heat the plasma significantly more effectively than hydrogen beams. Power balance analysis shows that in regimes with deuterium injection, ion heat diffusivity is close to neoclassical values (see Figure 8), even though the plasma remains unstable to ITG-mode development (see Figure 9). Deuterium injection also allow to reach significantly higher toroidal rotation velocities (see Figure 7a) and, consequently, stronger E×B shear, that is favorable for ITG turbulence suppressesion. As a result, ion thermal confinement improves, enabling ion heating up to 4 keV.

These findings suggest that the dominant factor influencing energy confinement time (see Figure 8c) is not the average mass of the main plasma ion (*M*) (how it was suggested in [23]), but rather the mass of the injected atom and the associated higher toroidal momentum input. In our experiments there were very small difference in shots with deuterium and hydrogen injection: M=1.6 and M=1.8 correspondently according to NPA measurements. However, the main ion mass may still appear to affect electron thermal confinement to some extent. As the fraction of deuterium in the plasma increases relative to hydrogen, electron heat diffusivity also tends to decrease (see Figure 8c).

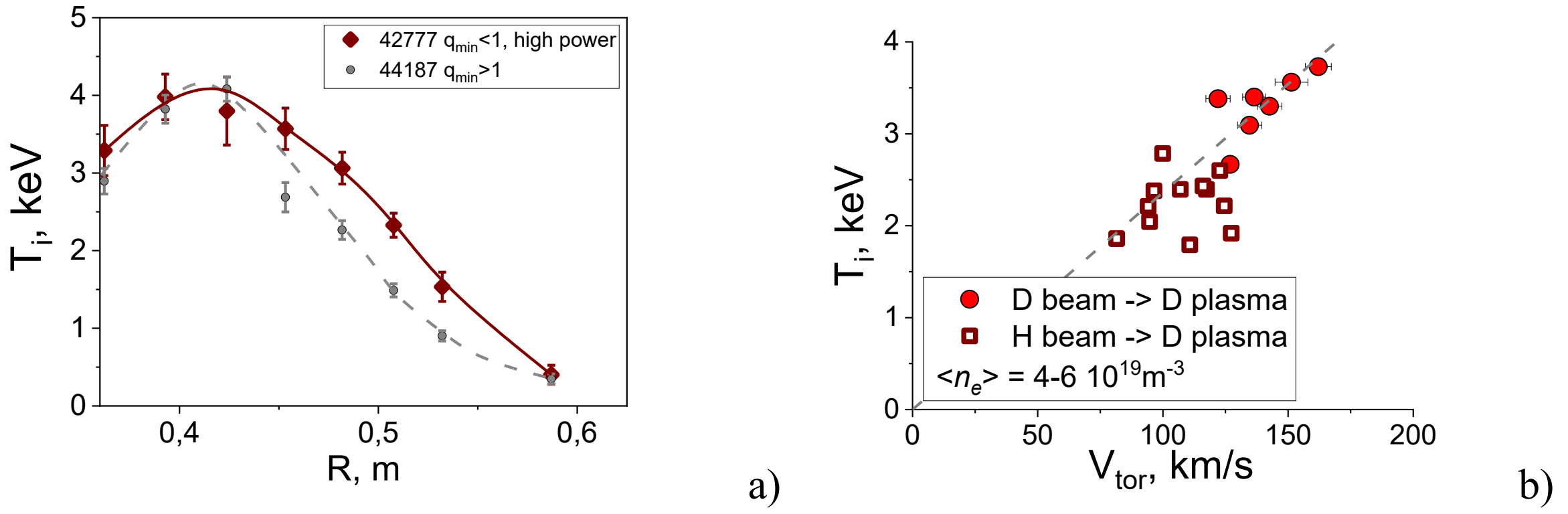

*Figure 5. (a) Ion temperature profile in the presence of sawtooth oscillations under double-beam heating; (b) Dependence of ion temperature on the toroidal rotation velocity at the plasma core for heating with high-energy hydrogen and deuterium neutral beams.*

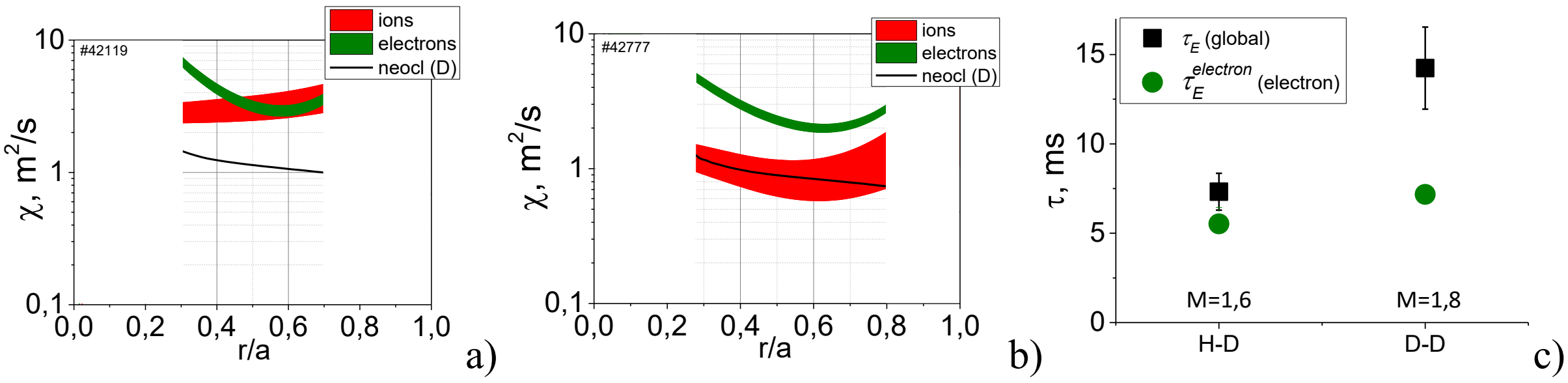

*Figure 6. Transport coefficients in hot ion mode during hydrogen beams heating; (b) Same for deuterium beams; (c) Comparison of total energy confinement time (black squares) and electron energy confinement time (green circles).*

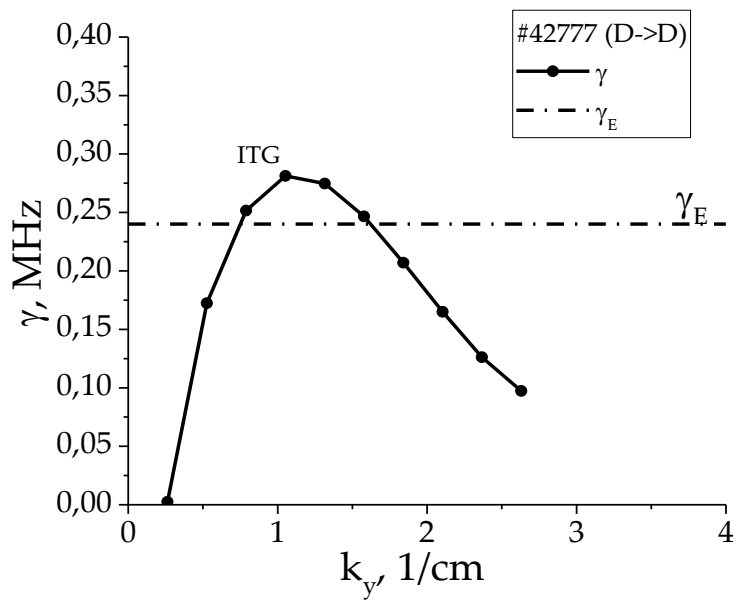

*Figure 7. ITG growth rates calculated for the gradient region (r/a = 0.7) compared with the estimated E×B shear in a discharge with $T_i$ = 4 keV heated by a deuterium neutral beam.*

## ***Discussion***

A hot ion mode was also achieved on the ST40 tokamak at toroidal magnetic fields of $B_T$ = 1.7–2 T [22,23]. Figure 10a presents a comparison of the H-factor versus toroidal field and heat diffusivity versus effective collisionality, combining data from both Globus-M2 and ST40. On Globus-M2, thermal confinement improves significantly with increasing toroidal magnetic field. However, this trend saturates at an H-factor (the ratio of energy confinement time to predictions of IPB98(y,2) scaling) of ~1.4, indicating a limit to confinement enhancement in spherical tokamaks with further $B_T$ increase. Despite a twofold difference in toroidal magnetic field between ST40 and Globus-M2, both ion and electron heat diffusivities are similar in magnitude, Figure 10b. This further supports the scaling saturation hypothesis. It emphasizes that predictive modeling using such kind of scalings of plasma parameters for future spherical tokamaks should be approached with caution. And such kind of simulations should be based on first-principles physics-based models.

Achieving good ion thermal confinement requires injecting sufficient toroidal momentum, which is more effectively delivered by heavier deuterium atoms. On ST40, switching from hydrogen to deuterium injection led to a clear improvement in energy confinement time, confirming the role of momentum input. However, a linear dependence between energy confinement time and the average mass of the main plasma ion was also observed, suggesting that the isotope effect also influences electron thermal confinement.

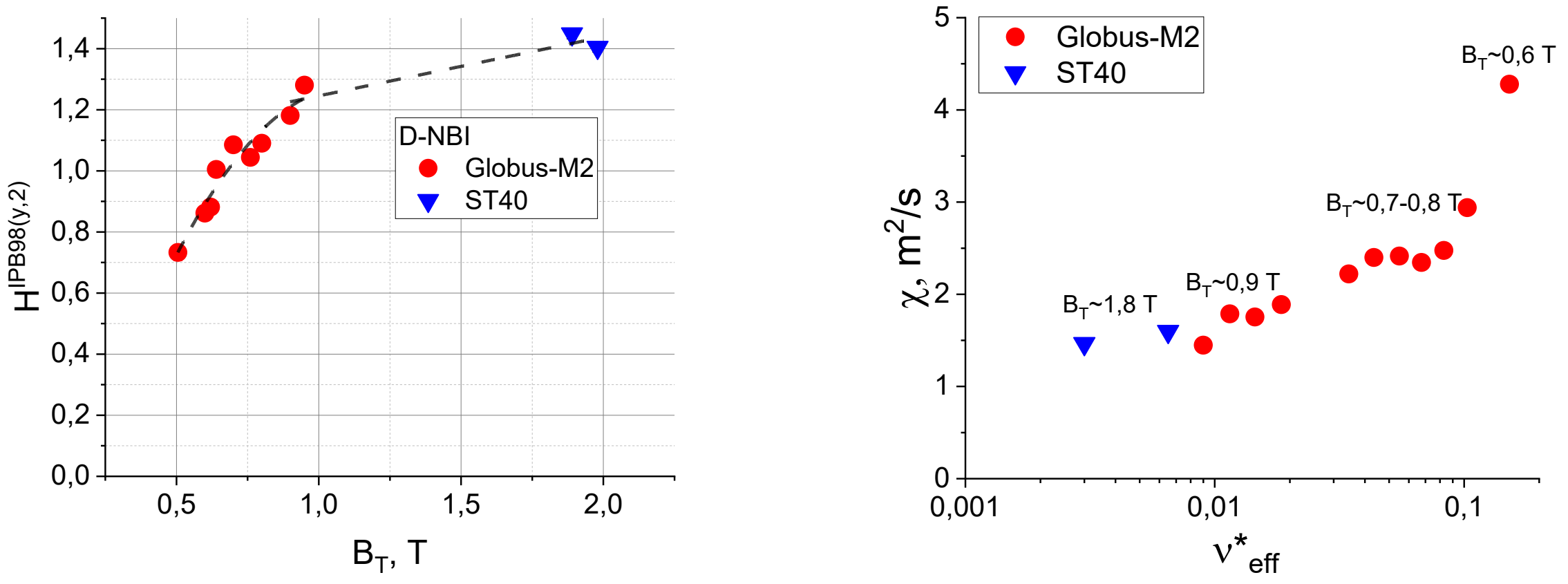

a) b)

*Figure 8. (a) Dependence of the H-factor on the toroidal magnetic field for Globus-M2 and ST40 data [23]. (b) average heat diffusivity vs collisionality for Globus-M2 and ST40 data [23].*

## ***Conclusion.***

On compact spherical tokamaks with $B_T \geq$ 0.7 T and neutral beam heating, the hot ion mode appears to be a natural operating regime. In contrast to recent results from conventional tokamaks such as KSTAR [24] where hot ion mode was achieved at relatively low plasma densities, spherical tokamaks can reach comparable ion temperatures at much higher plasma densities. On Globus-M2 the hot ion mode was successfully achieved even in a presence of sawtooth oscillations. However, in their absence, ion temperatures could be even higher. The main challenge for ion heating in spherical tokamaks may not be excessive energy transport, but rather MHD instabilities. These include resistive modes, which are typical for large-scale devices like MAST [25] and NSTX [26], and Alfven-type modes, which can be relevant under Globus-M2 conditions.

## ***Acknowlegments.***

The experiments were carried out at the Unique Scientific Facility "Spherical Tokamak Globus-M", which is incorporated in the Federal Joint Research Center "Material science and characterization in advanced technology". Plasma heating experiments with NBI presented in (section *Ion heating during plasma current ramp-up phase*) were financially supported by the project No. FFUG-2021-0001. The measurements of electron temperature and density spatial distributions (section *Ion heating during plasma current ramp-up phase*) were financially supported by the project No. FFUG-2024-0028. The energy confinement study (sections *Effect of* $I_p$ *and* $B_T$ *on* $\tau_E$ and *Effect of injected ion mass on ion heating*) and gyrokinetic simulations (section *Effect of injected ion mass on ion heating*) were financially supported by RSF #24-12-00162.

## Literature

1. M. Gryaznevich, R. Akers, P. G. Carolan et. al. "Achievement of Record b in the START Spherical Tokamak" Phys. Rev. Lett., 1988, 80, 3972 DOI: https://doi.org/10.1103/PhysRevLett.80.3972
2. J.E. Menard, M.G. Bell, R.E. Bell et. al. "β-Limiting MHD instabilities in improved-performance NSTX spherical torus plasmas" Nucl. Fusion 43 (2003) 330–340 **DOI** 10.1088/0029-5515/43/5/305
3. S M Kaye, J W Connor and C M Roach "Thermal confinement and transport in spherical tokamaks: a review" 2021 *Plasma Phys. Control. Fusion* **63** 123001**DOI** 10.1088/1361-6587/ac2b38
4. S.M. Kaye, M.G. Bell, R.E. Bell et. al. "Energy confinement scaling in the low aspect ratio National Spherical Torus Experiment (NSTX)" 2006 Nucl. Fusion 46 848 DOI 10.1088/0029-5515/46/10/002
5. R J Akers, J W Ahn, G Y Antar et. al. "Transport and confinement in the Mega Ampère Spherical Tokamak (MAST) plasma" 2003 Plasma Phys. Control. Fusion 45 A175 DOI 10.1088/0741-3335/45/12A/013
6. W. Guttenfelder, J. Candy, S. M. Kaye et. al. "Electromagnetic Transport from Microtearing Mode Turbulence" Phys. Rev. Lett. 106, 155004 DOI: https://doi.org/10.1103/PhysRevLett.106.155004
7. E. Mazzucato, D. R. Smith, R. E. Bell et. al. "Short-Scale Turbulent Fluctuations Driven by the Electron-Temperature Gradient in the National Spherical Torus Experiment" Phys. Rev. Lett. 101, 075001 DOI: https://doi.org/10.1103/PhysRevLett.101.075001

8. Y. Ren, S. M. Kaye, E. Mazzucato et. al. “Density Gradient Stabilization of Electron Temperature Gradient Driven Turbulence in a Spherical Tokamak” Phys. Rev. Lett. 106, DOI: https://doi.org/10.1103/PhysRevLett.106.165005
9. M. Valovič, R. Akers, M. de Bock et. al. “Collisionality and safety factor scalings of H-mode energy transport in the MAST spherical tokamak” 2011 Nucl. Fusion 51 073045 DOI 10.1088/0029-5515/51/7/073045
10. S.M. Kaye, S. Gerhardt, W. Guttenfelder et. al. “The dependence of H-mode energy confinement and transport on collisionality in NSTX” 2013 Nucl. Fusion 53 063005 DOI 10.1088/0029-5515/53/6/063005
11. G.S. Kurskiev, V.K. Gusev, N.V. Sakharov et. al. “Energy confinement in the spherical tokamak Globus-M2 with a toroidal magnetic field reaching 0.8 T” G.S. Kurskiev et al 2022 Nucl. Fusion 62 016011 DOI 10.1088/1741-4326/ac38c9
12. G.S. Kurskiev, I.V. Miroshnikov, N.V. Sakharov et. al. “The first observation of the hot ion mode at the Globus-M2 spherical tokamak” 2022 Nucl. Fusion 62 104002 DOI 10.1088/1741-4326/ac881d
13. G. S. Kurskiev, V. B. Minaev, N. V. Sakharov et. al. “Confinement, heating, and current drive study in Globus-M2 toward a future step of spherical tokamak program in Ioffe Institute” Phys. Plasmas 31, 062511 (2024) https://doi.org/10.1063/5.0211866
14. H. Anand, O. Bardsley, D. Humphreys et. al. “Modelling, design and simulation of plasma magnetic control for the Spherical Tokamak for Energy Production (STEP)” Fusion Engineering and Design, Volume 194, September 2023, 113724 https://doi.org/10.1016/j.fusengdes.2023.113724
15. Yunfeng LIANG, Huasheng XIE, Yuejiang SHI et. al. “Overview of the physics design of the EHL-2 spherical torus” 2025 Plasma Sci. Technol. 27 024001 DOI 10.1088/2058-6272/ad981a
16. V. B. Minaev, A. B. Mineev, N. V. Sakharov “Development of Next-Generation Spherical Tokamak Concept. The Globus-3 Tokamak” Plasma Phys. Rep. 49, 1578–1587 (2023). https://doi.org/10.1134/S1063780X23601189
17. G.S. Kurskiev, V.K. Gusev, S.Yu. Tolstyakov et. al. «Formation of Internal Transport Barriers in Globus-M Tokamak in Regime with Early Neutral Heating Beam Switch-On» Technical Physics Letters, 2011, Vol. 37, No. 12, pp. 1127–1131 DOI: 10.1134/S106378501112008X
18. A. Yu. Telnova, G. S. Kurskiev, E.O. Kiselev et. al. « Influence of the safety factor profile on the particle and heat transport in the Globus-M spherical tokamak» 2019 Plasma Sci. Technol. 21 115101 DOI 10.1088/2058-6272/ab2ff6
19. Jenko F and Dorland W “Prediction of Significant Tokamak Turbulence at Electron Gyroradius Scales” 2002 Phys. Rev. Lett. 89 225001 https://doi.org/10.1103/PhysRevLett.89.225001
20. Peeters A G, C. Angioni; M. Apostoliceanu et. al. “Linear gyrokinetic stability calculations of electron heat dominated plasmas in ASDEX Upgrade” 2005 Phys. Plasmas 12 022505 https://doi.org/10.1063/1.1848111
21. G.S. Kurskiev, N.N. Bakharev, V.V. Bulanin et. al. “Thermal energy confinement at the Globus-M spherical tokamak”, Nucl. Fusion 59 (2019) 066032 (7pp)? https://doi.org/10.1088/1741-4326/ab15c5
22. S.A.M. McNamara, O. Asunta, J. Bland et. al. “Achievement of ion temperatures in excess of 100 million degrees Kelvin in the compact high-field spherical tokamak ST40” 2023 Nucl. Fusion 63 054002 DOI 10.1088/1741-4326/acbec8

23. S M Kaye, M Sertoli, P Buxton et. al. "Isotope dependence of transport in ST40 hot ion mode plasmas" 2023 Plasma Phys. Control. Fusion 65 095012 DOI 10.1088/1361-6587/ace849
24. H. Han, S. J. Park, C. Sung et. al. "A sustained high-temperature fusion plasma regime facilitated by fast ions" Nature volume 609, pages269–275 (2022) https://doi.org/10.1038/s41586-022-05008-1
25. I.T. Chapman, M.-D. Hua, S.D. Pinches et. al. "Saturated ideal modes in advanced tokamak regimes in MAST" Nucl. Fusion 50 (2010) 045007 (16pp) doi:10.1088/0029-5515/50/4/045007
26. J.E. Menard, R.E. Bell, E.D. Fredrickson et. al. "Internal kink mode dynamics in high-β NSTX plasmas" Nucl. Fusion 45 (2005) 539–556 doi:10.1088/0029-5515/45/7/001